\documentclass[a4paper,11pt]{article}

\usepackage{jheppub} 

\usepackage[T1]{fontenc} 
\usepackage{float}
\usepackage{amsmath,amssymb,graphics,epsfig,subfigure}
\usepackage{color}
\usepackage{hyperref}
\usepackage{epstopdf}
\usepackage{graphicx}
\usepackage{braket} 
\usepackage{amsthm}
\usepackage{cases}

\newcommand \beq{\begin{equation}}
\newcommand \eeq{\end{equation}}
\newcommand \beqn{\begin{eqnarray}}
\newcommand \eeqn{\end{eqnarray}}
\newcommand \bseq{\begin{subequation}}
\newcommand \eseq {\end{subequations}}

\title{Generalized Volume-Complexity for RN-AdS Black Hole}

\author{Meng-Ting Wang,$^{a,b,c}$}
\author{Hong-Yue Jiang$^{a,b,c}$}
\author{and Yu-Xiao Liu$^{a,b,c}$
\footnote{Corresponding author}}

\affiliation[a]{Institute of Theoretical Physics $\&$ Research Center of Gravitation, Lanzhou University, Lanzhou 730000, China}
\affiliation[b]{Key Laboratory of Quantum Theory and Applications of MoE, Lanzhou University, Lanzhou 730000, China}
\affiliation[c]{Lanzhou Center for Theoretical Physics $\&$ Key Laboratory of Theoretical Physics of Gansu Province, Lanzhou University, Lanzhou 730000, China}

\emailAdd{wangmt21@lzu.edu.cn}
\emailAdd{jianghy21@lzu.edu.cn}
\emailAdd{liuyx@lzu.edu.cn}

\abstract
          { The connection between quantum information and quantum gravity has captured the imagination of physicists. Recently, a broad new class of gravitational observables have been proposed to provide new possibilities for holographic complexity \cite{Belin:2021bga} , which is an extension of volume in the Complexity=Volume proposal.
          In this paper, we investigate generalized volume-complexity for the 4-dimensional Reissner-Nordstr\"{o}m-AdS black hole. These new observables satisfy the characteristic of the thermofield double state, i.e., they grow linearly in time on the late stage.
          We find that there are multiple extremal hypersurfaces anchored at a certain boundary time. In other words, for the same boundary time, more than one observable (generalized volume-complexity) can exist in the bulk.
          The size relationship of the observables on the two hypersurfaces changes over time.  This will result in the substitution
          of the maximum extreme hypersurface which is dual to the complexity of the thermofield double state.
          We call the time when one hypersurface replaces another
            to become the largest extreme hypersurface the turning time $\tau _{turning}$.
            That is, a hypersurface dual to the complexity of the thermofield double state defined on the boundary jumps from one branch to another. This discontinuous jump is highly reminiscent of a phase transition, and the turning time denotes the moment at which this phase transition occurs.
            Our findings propose a discontinuous variation in bulk physics that is dual to the complexity of the thermofield double state defined on the boundary.}

\begin{document}
\maketitle
\flushbottom
\section{Introduction}

In 1997, Maldacena proposed the duality between a particular quantum gravity theory in $(d+1)$-dimensional asymptotically Anti-de Sitter (AdS) spacetime and a certain conformal field theory (CFT) living on the boundary of a $d$-dimensional spacetime \cite{Maldacena:1997re}. This duality is known as the AdS/CFT correspondence and provides the ground for kinds of conjectures about the connections between quantum information and quantum gravity \cite{Aaronson:2016vto,Nishioka:2009un,Chapman:2021jbh,Baggioli:2021xuv,OuYang:2020mpa,Faulkner:2022mlp}.

In 2006, Takayanagi proposed the conjecture of holographic entanglement entropy \cite{Ryu:2006bv}. It states that the entanglement entropy of a subregion on the boundary of a $d+1$-dimensional AdS space-time is equal to the minimum area of this subregion extending into the interior of the space-time. This conjecture implies a connection between the space-time geometry and the entanglement entropy. In 2013, Maldacena and Susskind proposed the conjecture that the Einstein-Rosen bridge and Einstein-Podolsky-Rosen entanglement may be considered as dually equivalent \cite{Maldacena:2013xja, Susskind:2016jjb}. It highlights the deep intrinsic connection between quantum
entanglement and space-time, which links general relativity to quantum mechanics,
and implies that quantum entanglement plays an important role in the fabric of space-time.
The volume of the wormhole for a two-sided AdS black hole grows linearly
in time on the late stage and continues growing far beyond
the time at which the entanglement entropy has thermalized \cite{Hartman:2013qma}.
Because entanglement entropy grows only for a short time, it is not sufficient to describe the volume growth of a wormhole by quantum entanglement alone \cite{Susskind:2014moa}. Thus, holographic complexity was proposed to describe the late volume growth of the Einstein-Rosen bridge, and a series of proposals were put forward.
The three most widely known hypotheses are ``complexity $=$ volume" (CV) \cite{Stanford:2014jda, Susskind:2014rva}, ``complexity $=$ action" (CA) \cite{Brown:2015bva, Brown:2015lvg}, and ``complexity $=$ spacetime volume" (CV2.0) \cite{Couch:2016exn}.
Refer to \cite{Omidi:2022whq,Chapman:2018lsv,Swingle:2017zcd,Fu:2018kcp,Auzzi:2018pbc,An:2019opz,An:2018dbz,Cai:2020wpc,Feng:2018sqm,Zhou:2021vsm,Zhou:2023nza,Peng:2018vbe,Meng:2018vtl,Huang:2016fks,Miao:2017quj,Akhavan:2019zax,Omidi:2020oit,Bravo-Gaete:2020lzs,Zolfi:2023bdp,Reynolds:2017lwq,Abad:2017cgl,Momeni:2017mmc,Maldacena:2001kr,Cai:2016xho,Guo:2017rul} 
for recent works on the CV and CA hypotheses.

The central idea of the CV proposal is that the complexity is dual to the volume1 of the extremal (maximal) hypersurface anchored at boundary times $t_{\text{R}}=t_{\text{L}}=\tau /2$ (consider symmetry):
\begin{equation}
  C_V=\text{max} \left[\frac{V}{G_N \ell _{\text{bulk}}}\right],
  \end{equation}
where $G_N$ is the gravitational constant and $\ell _{bulk}$ is a certain length scale to make the
holographic complexity dimensionless. In this paper, we select $\ell _{bulk}=L$, where $L$ is the AdS radius \cite{Couch:2018phr}.

In general, one can summarize the CV proposal in two steps. The first step is to select an extremal hypersurface in the bulk. The second step is
to evaluate the volume of the extremal hypersurface anchored on the boundary time slice $\Sigma _{CFT}$. From this perspective, one can construct an infinite class of new gravitational observables
on codimension-one surfaces, defined in terms of two scalar functions $F_{1}(g_{\mu \nu} ;X^{\mu })$ and $F_{2}(g_{\mu \nu} ;X^{\mu })$ \cite{Belin:2021bga}, which depends on the bulk
 metric $g_{\mu \nu}$ and the embedding functions $X^{\mu }(\sigma ^{a})$ of the hypersurfaces. The extremization procedure becomes
\begin{equation}
  \delta _X \left[\int_{\Sigma }  \,d^{d}\sigma  \sqrt{h} \,  F_{2}(g_{\mu \nu} ;X^{\mu })\right]=0,
\end{equation}
where $h$ is the determinant of the induced metric on the hypersurface in the bulk.
This condition identifies a special extreme hypersurface in the bulk which is called $\Sigma _{F_{2}}$. The evolution of this special extreme hypersurface over time is shown in Fig.~\ref{Penrose} for the Penrose diagram of the Reissner-Nordstr\"{o}m (RN)-AdS spacetime.
The second step is to calculate a special observable
on the hypersurface, which is a
diffeomorphism-invariant observable defined as
\begin{equation}\label{O1}
  O_{F_1 ,\Sigma _{F_2}}(\Sigma _{\text{CFT}})=\frac{1}{G_N L} \int_{\Sigma_{F_2} }  \,d^{d}\sigma  \sqrt{h}\, F_{1}(g_{\mu \nu} ;X^{\mu })=0.
  \end{equation}
The scalar functions $F_1$ and $F_2$ can be different. When we choose $F_1 = F_2 =1$, it turns into the CV proposal. For the sake of simplicity, we are just going to consider $F_1 = F_2$ in this paper. This means that the function representing the observables on the hypersurface coincides with the function of the extremization procedure.

As already noted in Ref.~\cite{Belin:2022xmt} that Eq.~(\ref{O1}) gives a broad new class of gravitational observables in the AdS black hole. And these observables show two important properties of the thermofield double state. That is, the growth rate of the observables in the late time is linear, and they reproduce the switchback effect \cite{Qu:2021ius,Qu:2022zwq,Roberts:2014isa, Susskind:2013aaa}.
The switchback effect of holographic complexity in the multiple-horizon black holes was investigated in Ref.~\cite{Jiang:2020bae}.

The aim of the paper is to explore the generalized volume-complexity for a RN-AdS black hole in four dimensions. We choose the scalar functionals
\begin{equation}
    F_1=F_2=1+\lambda L^4C^2, \label{scalarF1F2}
\end{equation}
where $C^2\equiv C_{\mu \nu \rho \sigma }C^{\mu \nu \rho \sigma }$ denotes the square of the Weyl tensor for the bulk spacetime, $\lambda$ is the dimensionless parameter. We will prove that for a non-extremal black hole, the generalized volume-complexity$C_{\text{gen}}$ has a linear growth at late times, which is independent of the selection of the free parameter $\lambda$. The linear growth rate is determined by the maxima of the effective potential $U(r)$ which is located between the two horizons.
We will focus on the case where the effective potential has two local maxima between two event horizons since the hypersurfaces corresponding to the peaks of the potential with different local maxima have different later growth rates and hence different generalized volume-complexities at late times.
The generalized volume-complexityon each hypersurface increases with time. So the size relationship of the two generalized volume-complexities on the two hypersurfaces changes over time. This will result in the substitution of the maximum extreme hypersurface which is dual to the complexity of the thermofield double state.
We call the time when one hypersurface replaces another to become the largest extreme hypersurface the turning time $\tau _{\text{turning}}$.
The substitution of the maximum extreme hypersurface proposes a discontinuous variation in bulk physics that is dual to the complexity of the thermofield double state defined on the boundary.
We will numerically analyze the effect of the charge-to-mass ratio $Q/M$ and the free parameters $\lambda$ of the RN-AdS black hole on the turning time.

The remainder of our paper is organized as follows.
In Sec.~\ref{sec2}, we solve the extreme conditions for these new codimension-one observables (\ref{O1}) in the case of the RN-AdS black hole.
In Sec.~\ref{sec3}, we prove that these new observables satisfy the characteristics of the thermofield double state, i.e., it grows linearly on the late stage.
In Sec.~\ref{sec4}, we numerically calculate the range of the charge-to-mass ratio $Q/M$ and the parameter $\lambda$ when more than one local maximum of the effective potential exists between the two horizons. Furthermore, we investigate the effect of the charge-to-mass ratio $Q/M$ and the free parameters $\lambda$ on the turning time.
Finally, the discussion and conclusion are given in Sec.~\ref{sec5}.

\section{The generalized volume-complexityfor the 4-dimensional charged AdS black hole}
\label{sec2}


In this section, we focus on the generalized volume-complexityfor the 4-dimensional RN-AdS black hole. The related works on the holographic complexity of the charged AdS black holes can refer to \cite{Cai:2017sjv,Ovgun:2018jbm,Carmi:2017jqz,Ma:2018hni,Goto:2018iay,Jiang:2021pzf,Jiang:2020spf}. The metric for the RN-AdS black hole is
\begin{eqnarray}\label{M1}
  ds^2&=&-f(r)dt^2+f(r)^{-1}dr^2+r^2d\Omega ^{2},\\
  f(r)&=&1-\frac{2M}{r}+\frac{Q^2}{r^2}+\frac{r^2}{L^2} \nonumber \\
  &=&\frac{(r-r_1)(r-r_2)(L^2 +r^2 +r_{1}^{2} +r_{2}^{2} +r_1 r_2 +r(r_1 +r_2))}{r^2 L^2},
  \end{eqnarray}
where $r_1$ and $r_2$ are respectively the inner and outer horizons of the RN-AdS black hole and $L$
is the AdS radius. For the sake of convenience in writing, we let $r_{1}^{2} +r_{2}^{2} +r_1 r_2 = \gamma$. The mass $M$ and charge $Q$ of the black hole can be expressed as
\begin{eqnarray}
  M&=&\frac{(r_1+r_2)(L^2+r_{1}^{2}+r_{2}^{2})}{2L^2},\\
    Q&=&\frac{\sqrt{ r_1 r_2 (L^2+r_{1}^{2}+r_1 r_2+r_{2}^{2})}}{L}.
    \end{eqnarray}

    Since we are concerned with the hypersurface extending from the asymptotic boundary through the event horizon of the black hole into the interior,
so we use the infalling Eddington-Finkelstein coordinates to rewrite the metric~(\ref{M1}):
\begin{equation}
  ds^2=-f(r)dv^2+2dvdr+r^2d\Omega ^{2},
  \end{equation}
where $v=t+r_{*}(r)$ with $r_{*}(r)=-\int_{r }^{\infty}  \,f(r')^{-1}dr' $.

\begin{figure}
  \centering
  \includegraphics[width=15cm]{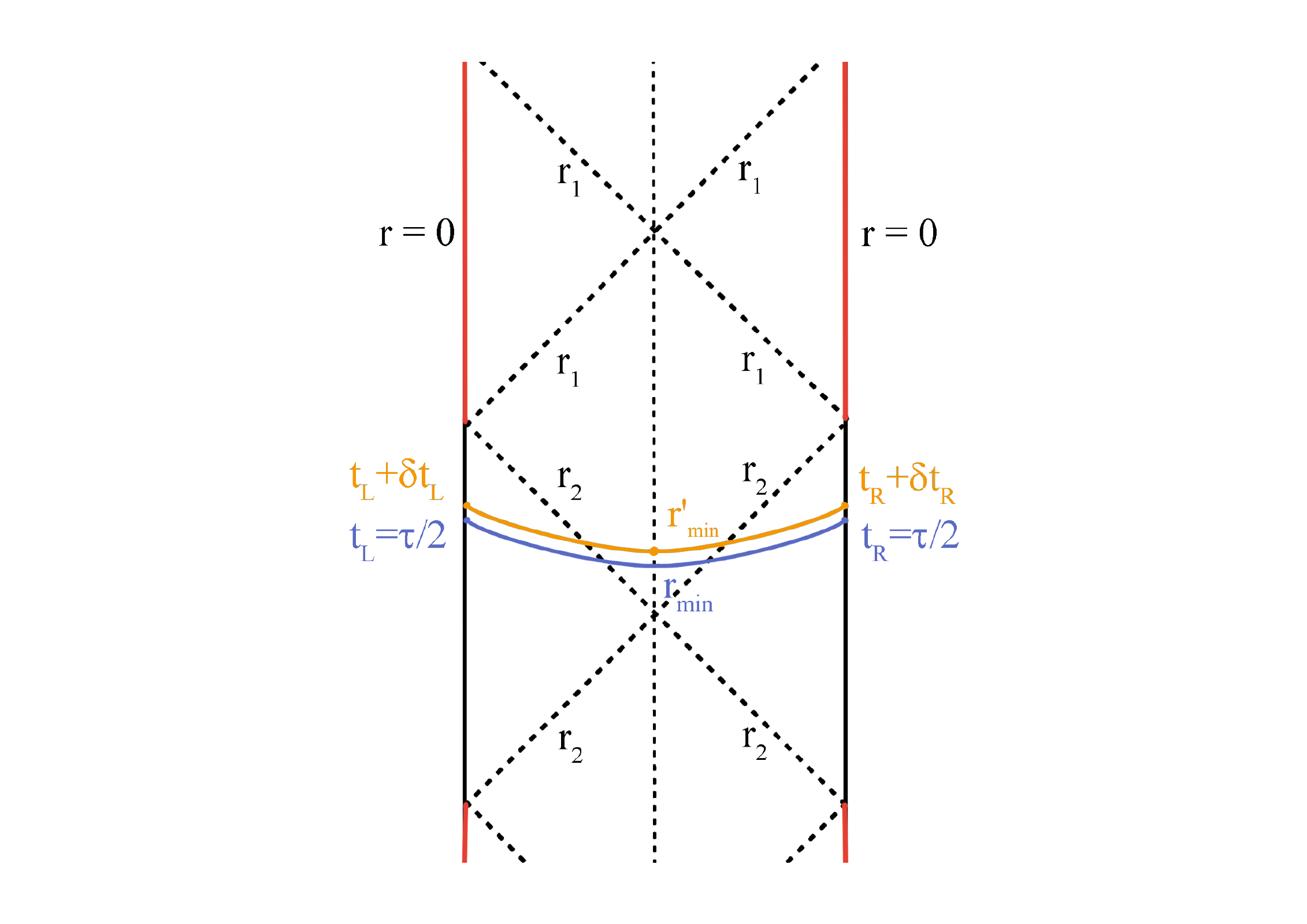}
  \caption{\label{Penrose}The Penrose diagram of the RN-AdS spacetime. When $t_{\text{L}}$ is shifted to $t_{L}+\delta t_{L}$, the extreme hypersurface evolves from the blue curve to the yellow curve.}
\end{figure}

The charged black hole geometry is dual to the charged thermofield double state
    \begin{equation}\label{ctfd}
      \ket{\psi _{\mathrm{cTFD}}(t_{\text{R}},t_{\text{L}})}=\sum_{n, \alpha}e^{-\beta (E_n - \mu Q_{\alpha })/2-iE_n (t_{\text{L}}+t_{\text{R}})}\ket{n, \alpha }_L \otimes \ket{n, \alpha }_R,
      \end{equation}
where $\mu $ is the chemical potential, $\beta$ is the inverse of the temperature, $E_n$ is the energy, $Q_{\alpha}$ is the charge, $L$ and $R$ label the quantum states ($\ket{n, \alpha}_L, \ket{n, \alpha}_R$), and the times ($t_{\text{L}},t_{\text{R}}$) are associated with the left and right
boundaries. To simplify, we set $t_{\text{L}}=t_{\text{R}}=\tau/2$, as shown in Fig.~\ref{Penrose}.

We choose the scalar functionals given in Eq.~(\ref{scalarF1F2}) and rewrite them again here
\begin{equation}\label{anything1}
  a(r)\equiv F_1=F_2=1+\lambda L^4C^2.
  \end{equation}
So the codimension-one observable becomes a covariant functional of the background curvature that can be evaluated on a codimension-one surface.
This observable can be seen as a generalization of the volume-complexity duality and, therefore, is referred to as the generalized volume-complexity$C_{\text{gen}}$.
Expressed as a functional parameterized by $(v(\sigma),r(\sigma),\Omega(\theta ,\phi ))$, the generalized volume-complexityfunctional can be represented as follows:
\begin{equation}\label{C0}
  C_{\text{gen}}=\frac{1}{G_N L}\int_{\Sigma _{F_{2}}  }  \,d^3\sigma \, \sqrt{h} \, a(r).
  \end{equation}
This expression can be further rewritten as
\begin{equation}\label{C1}
  C_{\text{gen}}=\frac{V_0}{G_N L}\int_{\Sigma _{F_{2}} }  \,d\sigma \, r^{2} \sqrt{-f(r) \dot v^{2} + 2\dot v \dot r} \, a(r),
  \end{equation}
  where $\sigma $ can be understood as a radial coordinate on
the hypersurface $\Sigma _{F_{2}}$, $V_0$ comes from the integral of the angular part. The point here is the derivative with respect to $\sigma $.
The generalized volume-complexity$C_{\text{gen}}$ is invariant under the transformation $\sigma \rightarrow \sigma'(\sigma)$, thus we can choose a reparameterized condition that is easier to calculate in the later calculations.

Since $C_{\text{gen}}$ is translationally invariant in the coordinate $v$, we can obtain the conserved quantity
\begin{equation}
  \frac{\partial L}{\partial \dot v }=r^{2}\frac{(\dot r -f(r)\dot v )a(r)}{\sqrt{-f(r) \dot v^{2} + 2\dot v \dot r}}\equiv P_v.
  \end{equation}
Choosing the parameter $\sigma$ as
  \begin{equation}\label{S1}
    \sqrt{-f(r) \dot v^{2} + 2\dot v \dot r}=a(r)r^{2},
    \end{equation}
we obtain the conserved momentum conjugate to $v$
\begin{equation} \label{P1}
  P_v=\dot r-f(r)\dot v.
  \end{equation}
Combining Eqs.~(\ref{S1}) and (\ref{P1}), we have the extremality conditions
\begin{eqnarray}\label{dott}
  \dot r&=&\pm \sqrt{P_{v}^{2}+f(r)a^2(r) r^{4}},\\
  \dot t&=&\dot v-\frac{\dot r}{f(r)}=\frac{-P_v \dot r}{f(r) \sqrt{P_{v}^{2}+f(r)a^2(r) r^{4}}}.
\end{eqnarray}
We can recast this problem as the motion of a classical non-relativistic particle in a potential.
The equation of motion reads as
\begin{equation}
  \dot r^2 +U(r)=P_{v}^{2},\label{R1}
\end{equation}
where the effective potential is
\begin{equation}
  U(r)=-f(r)a^2(r)r^{4}.
\end{equation}

Now, let us show that the effective potential of the four-dimensional RN-AdS black hole always has at least one local maximum between the inner horizon and outer horizons.
Using the metric~(\ref{M1}), we can get the square of the Weyl tensor as
\begin{align}
  C^2 & \equiv C_{\mu \nu \rho \sigma }C^{\mu \nu \rho \sigma } \\
 &=\frac{4\Big[-3r(r_1 +r_2)(r_1^2 +r_2^2)+6r_1 r_2 \gamma +L^2\big(r^2 +6r_1 r_2 -3r (r_1+r_2)\big)\Big]^2}{3L^4 r^8}.\notag
\end{align}
Because the effective potential vanishes at the two horizons of the black hole,
we only need to determine the sign of the derivative near the horizons.
The Cauchy principal values of the effective potential's first derivatives at the inner and outer horizons are respectively
\begin{eqnarray}
  \lim_{r \to r_1}  \frac{d U}{d r} &=&
       + \frac{r_1^2(r_2-r_1)(L^2 +3 r_1^2 +2r_1 r_2 +r_2^2) C_{0r_1}^2}{L^2}+\mathcal{O}(r-r_1) ,\\
  \lim_{r \to r_2}  \frac{d U}{d r} &=&
       -\frac{r_2^2(r_2-r_1)(L^2+r_1^2+2 r_1 r_2 +3 r_2^2) C_{0r_2}^2}{L^2}+\mathcal{O}(r-r_2),
  \end{eqnarray}
where $C_{0r_1}$ and $C_{0r_2}$ are expressions of $r_1$, $r_2$, and $L$. So, one can see that the first derivative of the effective potential is positive at the Cauchy horizon and negative at the event horizon, which implies there is at least one local maximum of the effective potential between these two horizons.


It is easy to verify that, for a RN-AdS black hole, the effective potential of the black hole vanishes at the two horizons.
For any given choice of the parameters $\lambda$, there must be at least one local maximum between the two horizons.
Furthermore, by adjusting the parameter $\lambda$ and the charge-to-mass ratio $Q/M$, multiple local maxima can be obtained between the two horizons.
The peak values of the effective potential near the Cauchy horizon and the event horizon are denoted by $U(r_{fL})$ and $U(r_{fR})$, respectively.
Here, $U(r_{fL})\equiv P_{\text{crtL}}^{2}$ and $U(r_{fR})\equiv P_{\text{crtR}}^{2}$ with $P_{\text{crtL},\text{crtR}}$ the critical momenta.
The local maximum of the effective potential occurs at $r=r_f$. We depict this behavior in Fig.~\ref{figur}.

  \section{The growth rate of the generalized volume}
  \label{sec3}

  Using Eqs.~(\ref{S1}) and (\ref{R1}), the functional $C_{\text{gen}}$ in Eq.~(\ref{C1}) can be written as follows:
  \begin{equation}\label{cg1}
    C_{\text{gen}}=2\frac{V_0}{G_N L}\int_{r_{\text{min}}}^{r_\epsilon }  \,dr \frac{r^{4} a^2(r)} {\sqrt{P_{v}^{2}-U(r)}},
    \end{equation}
where the $r_\epsilon$ is the UV cutoff and the $r_{\text{min}}$ is the turning point, which is the minimum radius on the extremal surface obtained by
\begin{equation}\label{u3}
  U(r_{\text{min}})=P_{v}^{2}.
  \end{equation}

  The boundary time can be obtained by the definition of the infalling coordinate:
    \begin{equation}\label{tb1}
      t_{\text{R}}+r^{*}(r_{\epsilon})-r^{*}(r_{\text{min}} )=\int_{r_{\text{min}}}^{r_\epsilon }  \,dr\frac{\dot \upsilon  }{\dot r}
      =\int_{r_{\text{min}}}^{r_\epsilon }  \,dr\frac{1}{f(r)}\left[1-\frac{P_v}{\sqrt{P_{v}^{2}+f(r)a^2(r) r^{4}}}\right].
    \end{equation}
So we can connect the boundary time $\tau$ to the conserved momentum:
     \begin{equation}\label{t1}
      \tau=2t_{\text{R}}=-2\int^{r_\epsilon }_{r_{\text{min}}}  \,dr\frac{P_v}{f(r)\sqrt{P_{v}^{2}+f(r)a^2(r) r^{4}}}.
    \end{equation}
The functional $C_{\text{gen}}$ can be further written as follows:
\begin{equation}\label{C2}
  \begin{split}
C_{\text{gen}}&=\frac{2 V_0}{G_N L} \int^{r_\epsilon }_{r_{\text{min}}} \,dr\frac{1}{f(r)}\left[\sqrt{P_{v}^{2}+f(r)a^2(r) r^{4}}-P_v\right]\\
&+\frac{2 V_0 P_v}{G_N L} \left(t_{\text{R}}+r^{*}(r_{\epsilon})-r^{*}(r_{\text{min}} )\right) .
  \end{split}
\end{equation}
Taking the time derivative of Eq.~(\ref{C2}) and using the Eq.~(\ref{tb1}), the time derivative
      of the generalized volume-complexityat the boundary time $\tau $ is given by
      \begin{equation}
        \frac{dC_{\text{gen}}}{d\tau }=\frac{V_0}{G_N L}P_v(\tau ).
        \end{equation}
The linear growth at late times becomes
        \begin{equation}
      \lim_{\tau \to \infty}   \frac{dC_{\text{gen}}}{d\tau} \sim  \lim_{\tau \to \infty} P_v(\tau )= P_{\text{crt}}.
          \end{equation}
This shows that the new observable we have constructed satisfies the basic characteristic of the constant growth rate of the thermofield double state at late times.

The relation between the boundary time $\tau$ and conserved momentum $P_v$ can be determined numerically by integrating Eq.~(\ref{t1}). It can be further written as
\begin{equation}\label{t2}
  \tau =2t_{\text{R}}=-2\int_{r_\epsilon }^{r_{\text{min}}} dr \frac{P_v}{f(r)\sqrt{P^{2}_{v}-U(r)} }.
  \end{equation}
In Eq.~(\ref{t2}), we can see two divergences. When the integral is near the horizon, $f(r) \to 0$. Further, when the integral is near the $r_{\text{min}}$, $\sqrt{P^{2}_{v}-U(r)} \rightarrow 0$.
However, the Cauchy principal value of this integral for the singularities at the horizon and $r_{\text{min}}$ is still finite. Hence, it does not affect the calculation of the boundary time.

  \section{Effective potential with two local maximum and turning time}
\label{sec4}

It is evident that the effective potential of the four-dimensional RN-AdS black hole vanishes at the horizon and there must be at least one local maximum between the two horizons.
The turning point $r_{\text{min}}$ determined by Eq.~(\ref{u3}) can be obtained from the following relation:
  \begin{equation}
    \begin{split}
&\frac{ r_{\text{min}}^{2}(r_1-r_{\text{min}})(r_{\text{min}}-r_2)(L^2 +r_{\text{min}}^2 +r_{1}^{2} +r_{2}^{2} +r_1 r_2 +r_{\text{min}}(r_1 +r_2))}{ L^2} \\
&= \frac{P_{v}^{2}}{(1+\lambda L^4C^2)^2}.
    \end{split}
    \end{equation}

  \begin{figure}
   \centering
   \includegraphics[width=10cm]{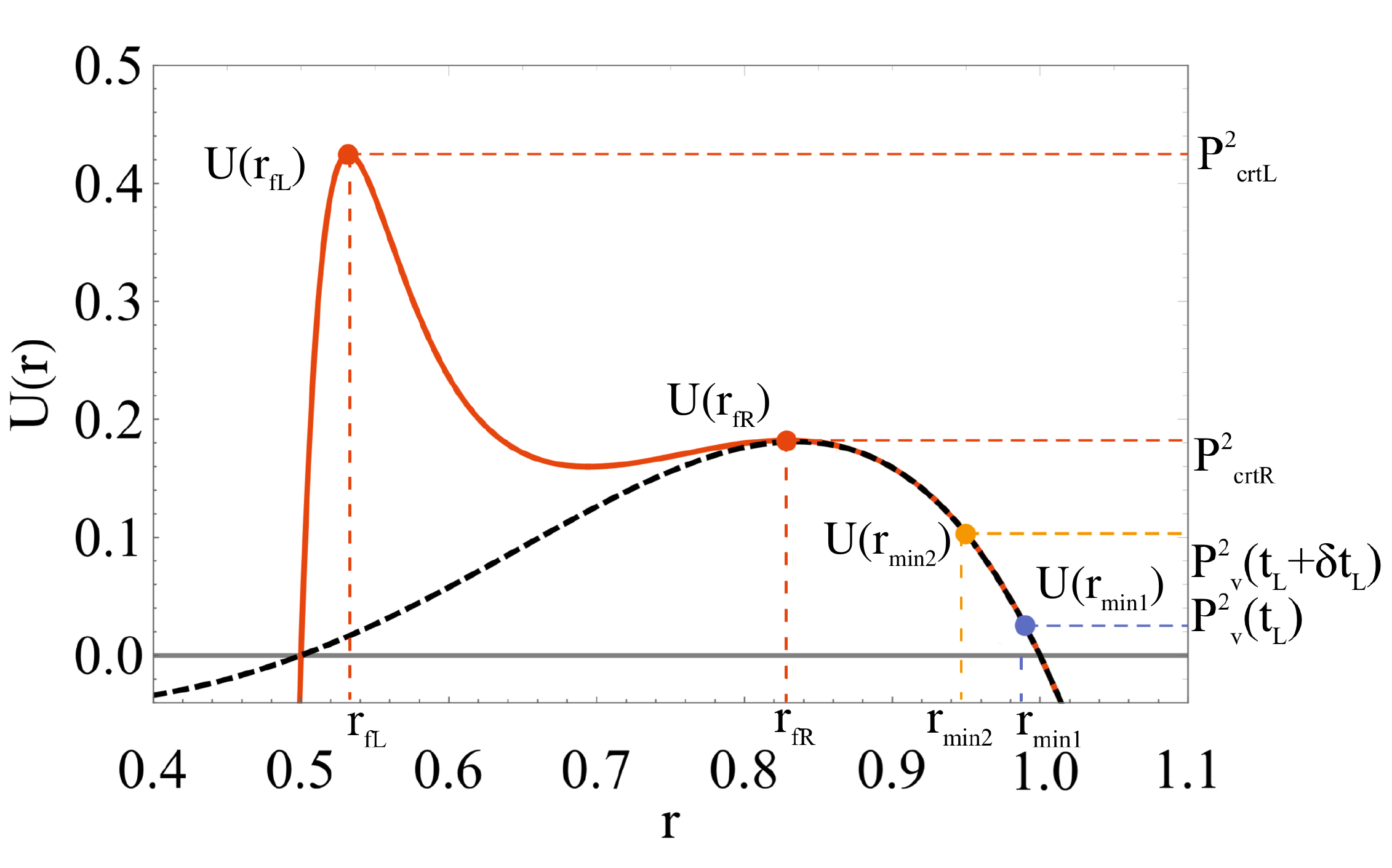}
   \caption{\label{figur}The effective potential $U$ as a function of the radial coordinate with $r_1/r_2=0.5$, $L/r_2=1$, and $\lambda=0.002$ and $0.0$. When $\lambda=0$, the generalized volume-complexitydegenerates to the volume $V$. The black dashed curve represents the potential obtained from the volume.}
  \end{figure}

\begin{figure}  
  \subfigure[~$\frac{Q}{M}-\lambda$]{\label{Phase}
    \includegraphics[width=7cm]{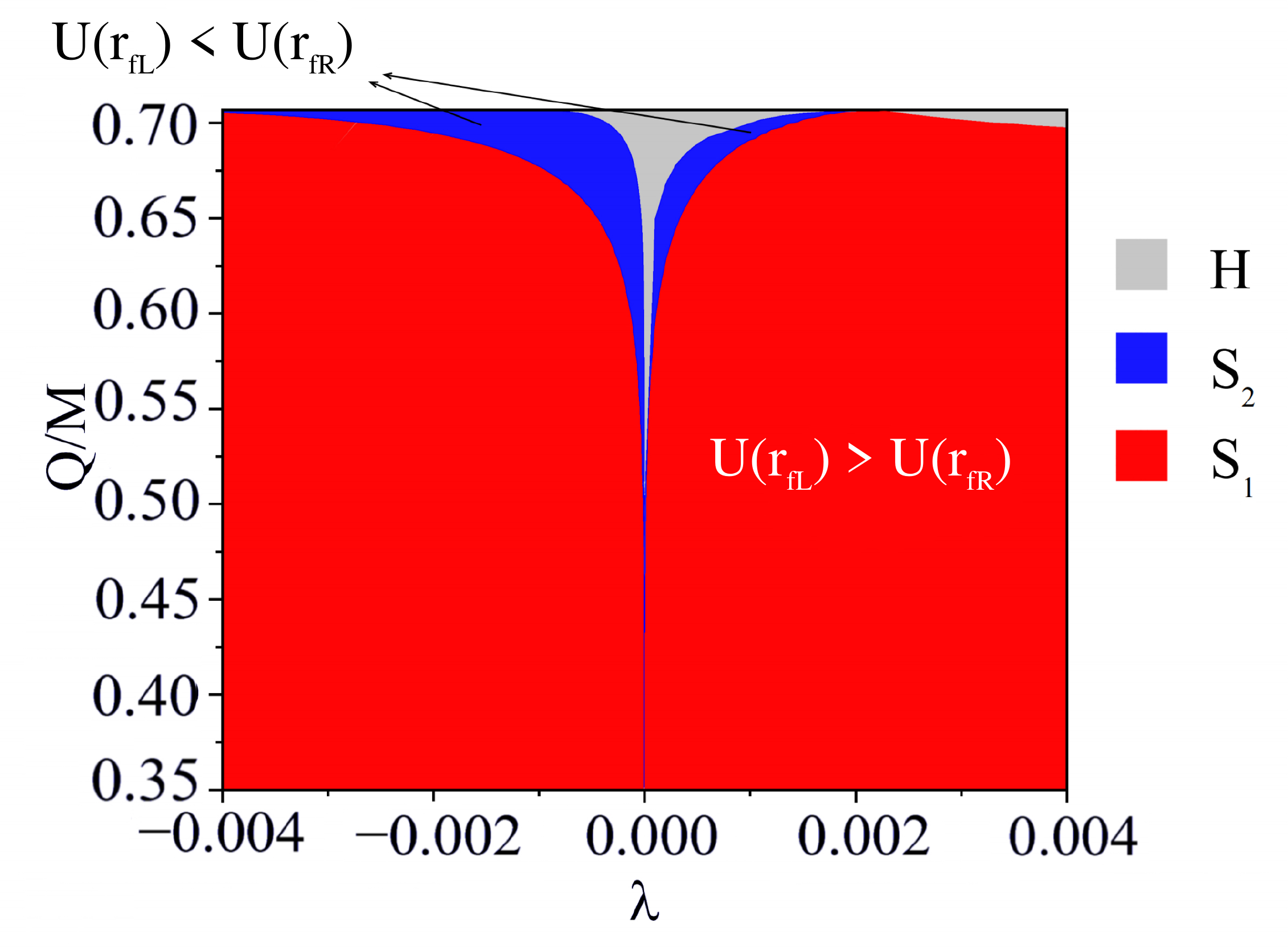}}
  \subfigure[~$U-r/r_2$]{\label{phaseur}
    \includegraphics[width=7cm]{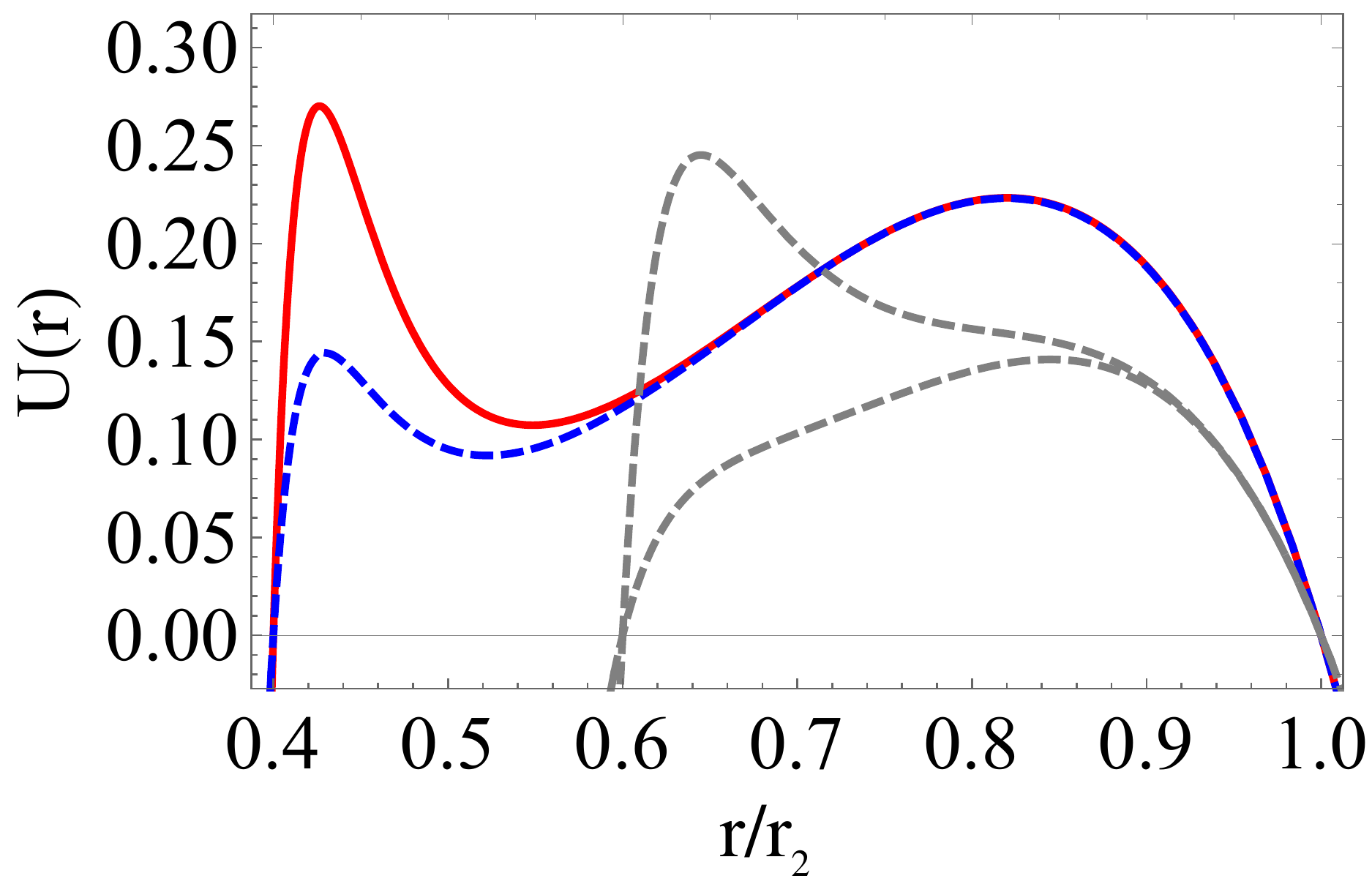}}
    \caption{Left: The parameter space of the number of the local maxima of the effective potential $U(r)$ between the two horizons of the black hole with $L/r_2=1$. In subspaces $S_1$ and $S_2$, there are two local maxima, $U(r_{fL})$ and $U(r_{fR})$. In subspace $S_1$, $U(r_{fL})$ is greater than $U(r_{fR})$, and in subspace $S_2$, $U(r_{fL})$ is smaller than $U(r_{fR})$. In subspace $H$, there is only one local maximum of the $U(r)$.
 Right: The effective potential $U(r)$ with $L/r_2=1$. The red curve corresponds to $r_1/r_2=0.4$ and $\lambda=0.0006$. The blue dashed curve corresponds to $r_1/r_2=0.4$ and $\lambda=0.0004$. The grey dashed curves correspond to $r_1/r_2=0.6$ and $\lambda=0.001, 0.003$.}
  \end{figure}

We show the parameter space of the number of the local maxima of the effective potential $U(r)$ for the four-dimensional RN-AdS black hole between the two horizons in Fig.~\ref{Phase}.
We define the parameter spaces $H$  and $S$ as the set of parameters $(\lambda, Q/M)$ for which the effective potential has one and two local maxima between the two horizons, respectively. Within this parameter space $S$, there are two subspaces $S_1$ and $S_2$, where the peak values near the Cauchy horizon are higher and lower than that near the event horizon, respectively.
The effective potentials corresponding to the  parameter spaces $S_1$, $S_2$ and $H$ are depicted by the red curve, blue curve and grey curve in Fig.~\ref{phaseur}, respectively.
  The late-time linear growth rate is determined by the maximum of the effective potential $U(r)$, i.e., $U(r_f)=P_{\text{crt}}^{2}$ (see Fig.~\ref{figur}). Having two local maxima implies the existence of two critical momenta $P_{\text{crt}}$. This means that for an infinite boundary time $\tau \rightarrow \infty$, we obtain two different extremal hypersurfaces, and the minimum radii on the hypersurfaces are $r_{fL}$ and $r_{fR}$. The two extremal hypersurfaces have different late-time linear growth rates $P_{\text{crtL}}$ and $P_{\text{crtR}}$. The parameter space $S_1$ is of particular interest to us because the hypersurface corresponding to the peak near the Cauchy horizon has a greater late-time linear growth rate.

  In particular, in the case of a near-extreme RN-AdS black hole, when the inner and outer event horizons of the black hole are sufficiently close, the two local maxima of the effective potential between the horizons become indistinguishable. Moreover, when the free parameter $\lambda $ takes on a small value, the effective potential reduces to the CV proposal, resulting in only one local maximum between the two horizons. This scenario, where there is a single local maximum between the event horizons, falls outside the scope of our current discussion. This can be observed in the parameter space denoted as $H$


By integrating Eq.~(\ref{t1}), we can establish the relationship between the boundary time $\tau$ and conserved momentum $P_v$.
However, we observe that the integrand has divergences near the horizons. By performing a series expansion near the horizon, we find that the Cauchy principal value of the integrand behaves as:
  \begin{gather}
    \lim_{r \to r_1} \frac{P_v}{f(r)\sqrt{P^{2}_{v}-U(r)}}  \sim \frac{r_1^2}{(r_1-r_2)(r-r_1)},\\
    \lim_{r \to r_2} \frac{P_v}{f(r)\sqrt{P^{2}_{v}-U(r)}} \sim \frac{r_2^2}{(-r_1+r_2)(r-r_2)}.
  \end{gather}
Thus, the integrand near the horizons diverges to the order of negative one. Therefore, although the integrand diverges near horizons, the integral~(\ref{t1}) will give a finite value.
Given that we are interested in the case where the minimum radius on the hypersurface is between the horizons, we can focus our attention on the divergence at $r_2$ and ignore the divergence at $r_1$, since it does not affect the extremal hypersurface of interest.

Similarly, as the integral approaches $r_{\text{min}}$, $\sqrt{P^{2}_{v}-U(r)}$ approaches zero.
By expanding the leading-order around the minimum radius $r_{\text{min}}$:
  \begin{equation}
    \lim_{r \to r_{\text{min}}} \frac{1}{\sqrt{P^{2}_{v}-U(r)}}  \sim \frac{1}{\sqrt{U'(r_{\text{min}})(r_{\text{min}}-r)}},
  \end{equation}
  the integral becomes regular around the minimal radius $r_{\text{min}}$.

\begin{figure}
    \centering
    \includegraphics[width=7cm]{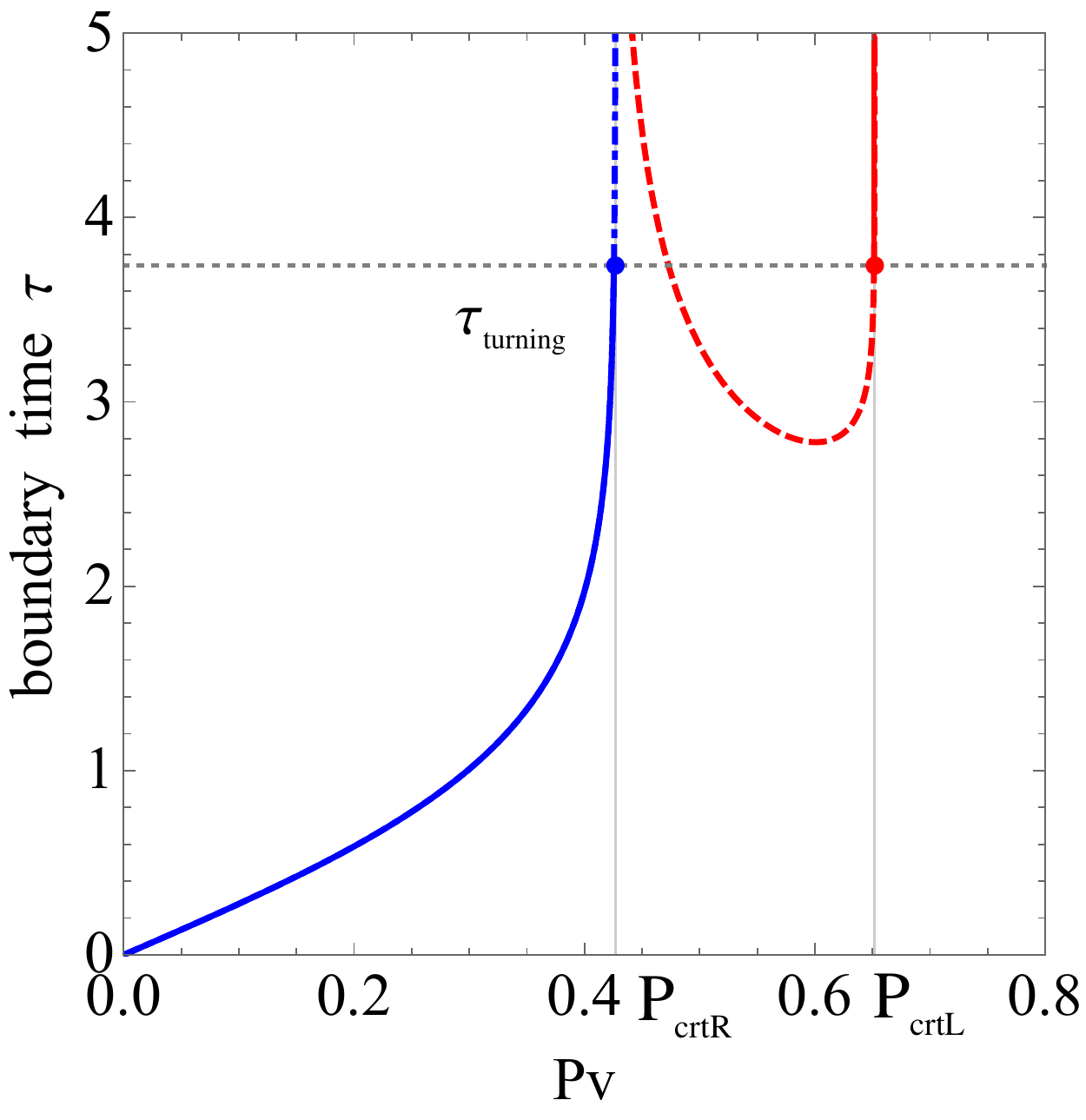}
    \caption{\label{tp1}The relation between the boundary time $\tau$ and the conserved momentum $P_v$ with $r_1/r_2=0.5$, $L/r_2=1$, $\lambda =0.002$.
    The two solid grey lines represent the critical momentum $P_{\text{crtL}}$ and $P_{\text{crtR}}$, respectively.
    The blue curve corresponds to hypersurfaces with $P_v < P_{\text{crtR}}$, and the minimum radius $r_{\text{min}}$ on the hypersurfaces satisfies $r_{\text{min}} > r_{fR}$ (see Fig.~\ref{figur}).
    The red curve corresponds to hypersurfaces where $P_{\text{crtR}} < P_v < P_{\text{crtL}}$, and the minimum radius $r_{\text{min}}$ on the hypersurfaces satisfies $r_{fL} < r_{\text{min}} < r_{fR}$.
    The solid lines are used to represent the
branches that represent the maximum generalized volume-complexityat different times.
The hypersurface corresponding to the blue branch has a larger generalized
volume before the turning time is reached. After the turning time, the hypersurface
corresponding to the red non-inclined branch has a larger generalized volume.}
   \end{figure}

The relationship between conserved momentum and the boundary time  is illustrated in Fig.~\ref{tp1}.
It should be noted that the red curve corresponds to two kinds of branches, which respectively correspond to two kinds of extremal hypersurfaces.
One approaches the solid grey line representing the critical momentum $P_{\text{crtR}}$ from the left,
and the other approaches the solid grey line representing the critical momentum $P_{\text{crtL}}$ from the right as time evolves.
As discussed in Ref.~\cite{Belin:2021bga}, the first type is known as the dipping branch and  is represented by the branch approaching the solid grey line from the left, which represents the critical momentum $P_{\text{crtR}}$.
The dipping branch does not correspond to the extremal surface with the maximal generalized volume-complexityat any time.


  Using Eqs.~(\ref{S1}) and (\ref{dott}), we can express the functional $C_{\text{gen}}$ in Eq.~(\ref{C1}) as:
  \begin{equation}
    \frac{G_N LC_{\text{gen}}}{2V_0}=\int_{t=0}^{t=t_{\text{R}}} \frac{U(r(t))}{P_{v}} \,dt.
    \end{equation}
Thus, we can then evaluate the late limit of the generalized volume-complexityof the non-dipping branch by
\begin{equation}
\frac{C_{\text{gen}} G_N L}{2V_0}=\int_{t=0}^{t \to \infty} P_{\text{crt}} \,dt.
  \end{equation}
Therefore, we obtain
\begin{equation}
  \frac{G_N L(C_{\text{genL}}-C_{\text{genR}})}{2V_0}=\int_{t=0}^{t \to \infty} (P_{\text{crtL}}-P_{\text{crtR}}) \,dt > 0,
  \end{equation}
where $C_{\text{genR}}$ represents the generalized volume-complexitycorresponding to the non-dipping branch of the blue curve where $P_{v}\to P_{\text{crtR}}$, and
$C_{\text{genL}}$ represents the generalized volume-complexitycorresponding to the non-dipping branch of the red curve where $P_{v}\to P_{\text{crtL}}$.
Thus, the extremal hypersurface corresponding to a larger late-time growth rate has a larger generalized volume. This implies that the extremal hypersurface with a higher effective potential peak near the Cauchy horizon, denoted as $U(r_{fL})$, has a larger generalized volume-complexitycompared to the extremal hypersurface with a lower effective potential peak near the event horizon, denoted as $U(r_{fR})$.

We use $C_{\text{genM}}$ to represent the generalized volume-complexitycorresponding to the dipping branch of the red curve where $P_{v}\to P_{\text{crtR}}$. Since $P_{\text{crtR}}^2 > U(r)$ near $r_{\text{fR}}$, we have
\begin{equation}
  \frac{G_N L(C_{\text{genR}}-C_{\text{genM}})}{2V_0}=\int_{t=0}^{t \to \infty} \left(P_{\text{crtR}}-\frac{U(r)}{P_{\text{crtR}}}\right) \,dt > 0,
  \end{equation}
thus,
\begin{equation}
  C_{\text{genL}}> C_{\text{genR}}>C_{\text{genM}}.
  \end{equation}

  Based on the previous results, we can introduce the concept of a turning time $\tau_{\text{turning}}$ in the evolution of the hypersurface.
  When the hypersurface evolves with the boundary time $\tau $, there will be a turning time $\tau_{\text{turning}}$. Before the turning time,
  the generalized volume-complexity$C_{\text{genL}}$ is smaller than $C_{\text{genR}}$, but after the turning time,
  the generalized volume-complexity$C_{\text{genL}}$ is larger than $C_{\text{genR}}$. Using  Eqs.~(\ref{t2}) and (\ref{cg1}), we can determine the turning time $\tau _{\text{turning}}$.

\begin{figure}
    \centering
    \includegraphics[width=7cm]{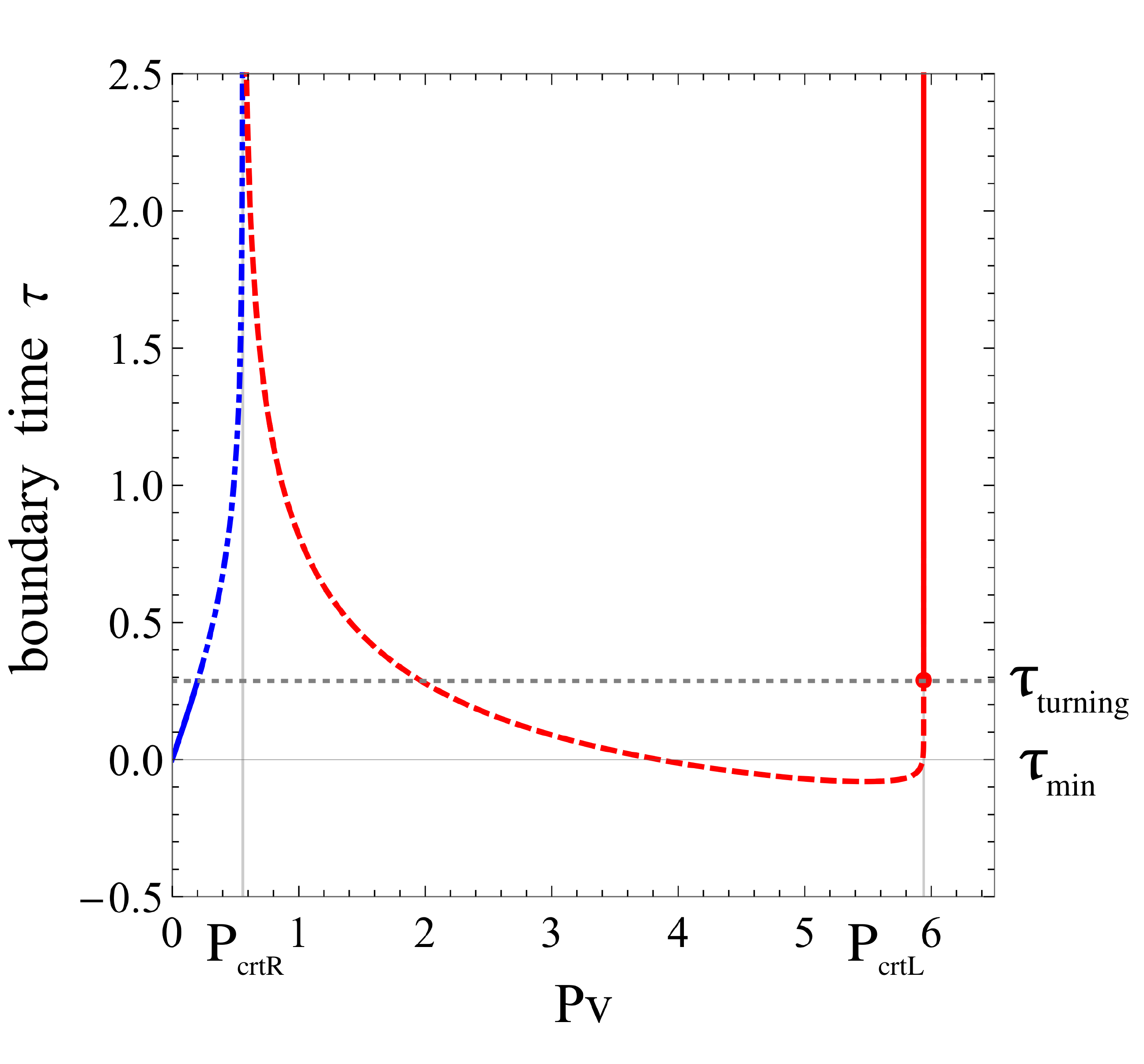}
    \caption{\label{tp2}The relation between the boundary time $\tau$ and the conserved momentum $P_v$ with $r_1/r_2=0.18$, $L/r_2=1$, and $\lambda =0.00025$.
    The two solid grey lines represent the critical momenta $P_{\text{crtL}}$ and $P_{\text{crtR}}$, respectively. The minimum negative boundary time and the turning time are denoted as $\tau_{\text{min}}$ and $\tau_{\text{turning}}$, respectively.}
   \end{figure}

\begin{figure}  
  \subfigure[~$\tau > \left| \tau_{\text{min}} \right| $]{\label{timereflecteda}
    \includegraphics[width=7cm]{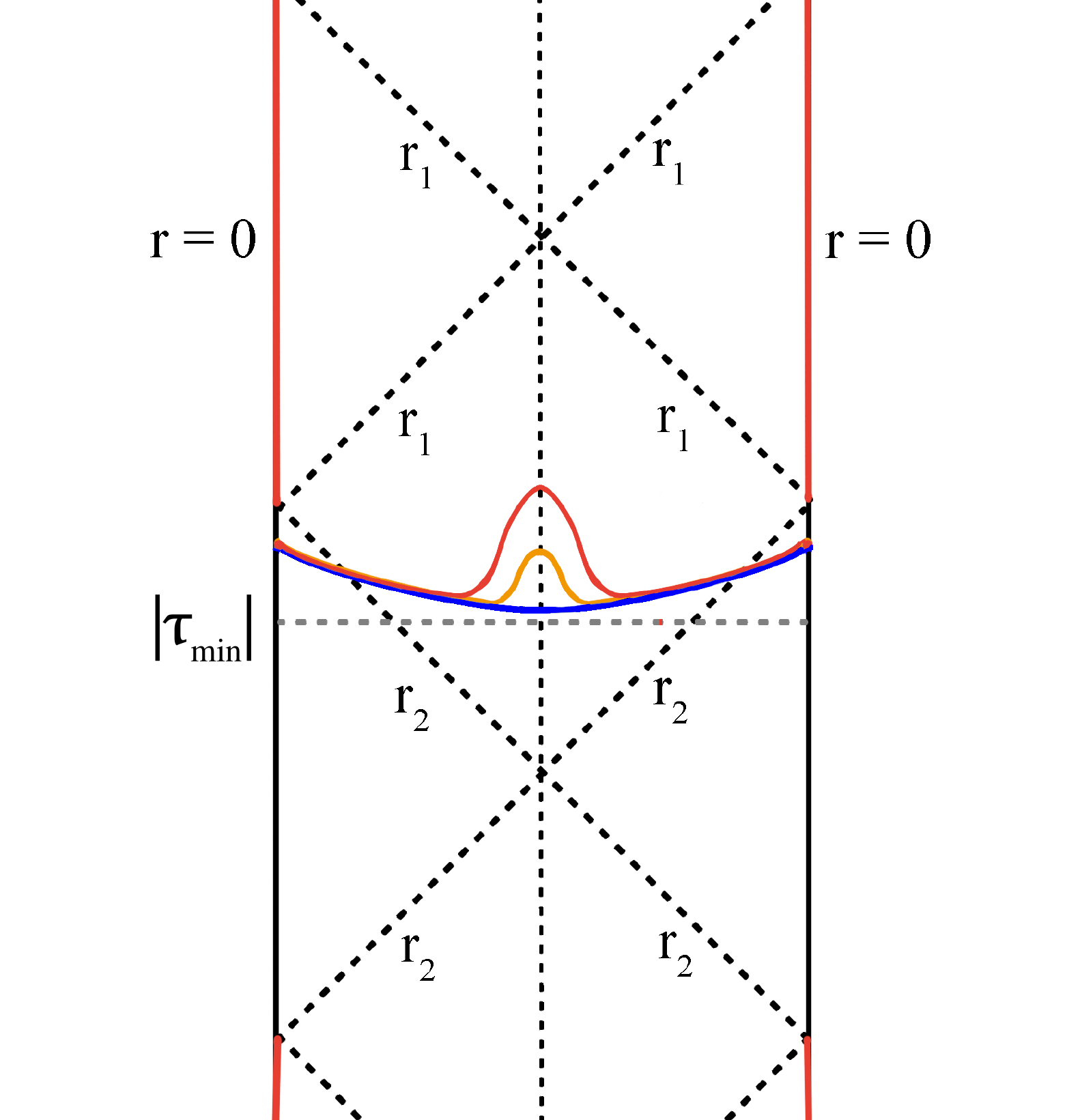}}
  \subfigure[~$\tau < \left| \tau_{\text{min}} \right| $]{\label{timereflectedb}
    \includegraphics[width=7cm]{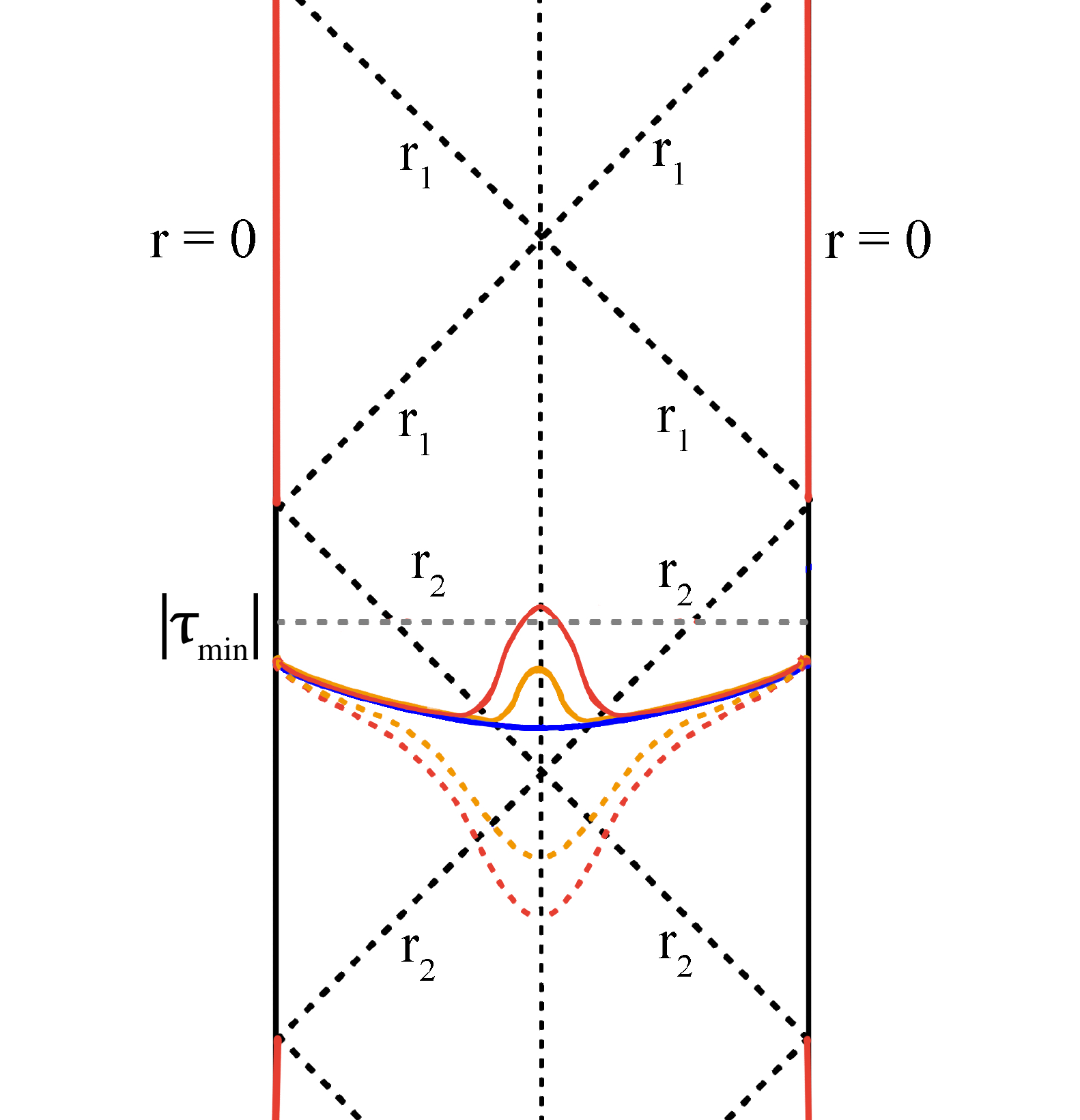}}
    \caption{Penrose diagram for the RN-AdS spacetime, Left $\tau > \left| \tau_{\text{min}} \right| $;
    Right: $\tau < \left| \tau_{\text{min}} \right| $.}
  \end{figure}

   We will first discuss the number of extreme surfaces for a given boundary time $\tau$.
   For the extremal hypersurface that has the minimum radius between the two horizons of the black hole,
   if the time to reach the boundary is always positive, as shown in Fig.~\ref{tp1}, we have three hypersurfaces for one boundary time $\tau$.
   However, when there is a negative part of the boundary time (see Fig.~\ref{tp2}), it needs to be discussed in different cases.
   When the boundary time is larger than the absolute value of the minimum negative boundary time $\tau > \left| \tau_{\text{min}} \right| $,
   we can obtain three extreme surfaces for the same boundary time, as shown in Fig.~\ref{timereflecteda}.
   Because of the reflection symmetry, when the boundary time is less than the absolute value of the minimum negative boundary time $\tau < \left| \tau_{\text{min}} \right| $,
   we can obtain five extreme surfaces for the same boundary time,
   as shown in Fig.~\ref{timereflectedb}, where the dotted line represents the time reflected trajectory.
   The question is which extremal hypersurface has the maximal volume.
   As we have shown before, the corresponding hypersurface
   that exhibits larger late growth has a larger general volume.

  The relation between the boundary time $\tau $ and generalized volume-complexity$C_{\text{gen}}$ can be numerically determined by integrating Eqs.~(\ref{cg1}) and (\ref{t2}), which are connected by the upper limit $r_{\text{min}}$ of the integration.

Before calculating the turning time, it is important to note that the magnitude of the relationship between the generalized volume
of two extremal hypersurfaces is independent of the selection of the
UV stage $r_{\epsilon }$. This can be demonstrated by performing a series expansion at $r \to \infty$, which reveals that the Cauchy principal value of the integrand is given by
\begin{equation}
  \lim_{r \to \infty} \big(C_{\text{genL}}-C_{\text{genR}}\big) \sim \frac{1}{r^2}.
\end{equation}
This shows that the difference between the generalized volume-complexities of the two hypersurfaces converges at infinity.

\begin{figure}
  \centering
  \includegraphics[width=7cm]{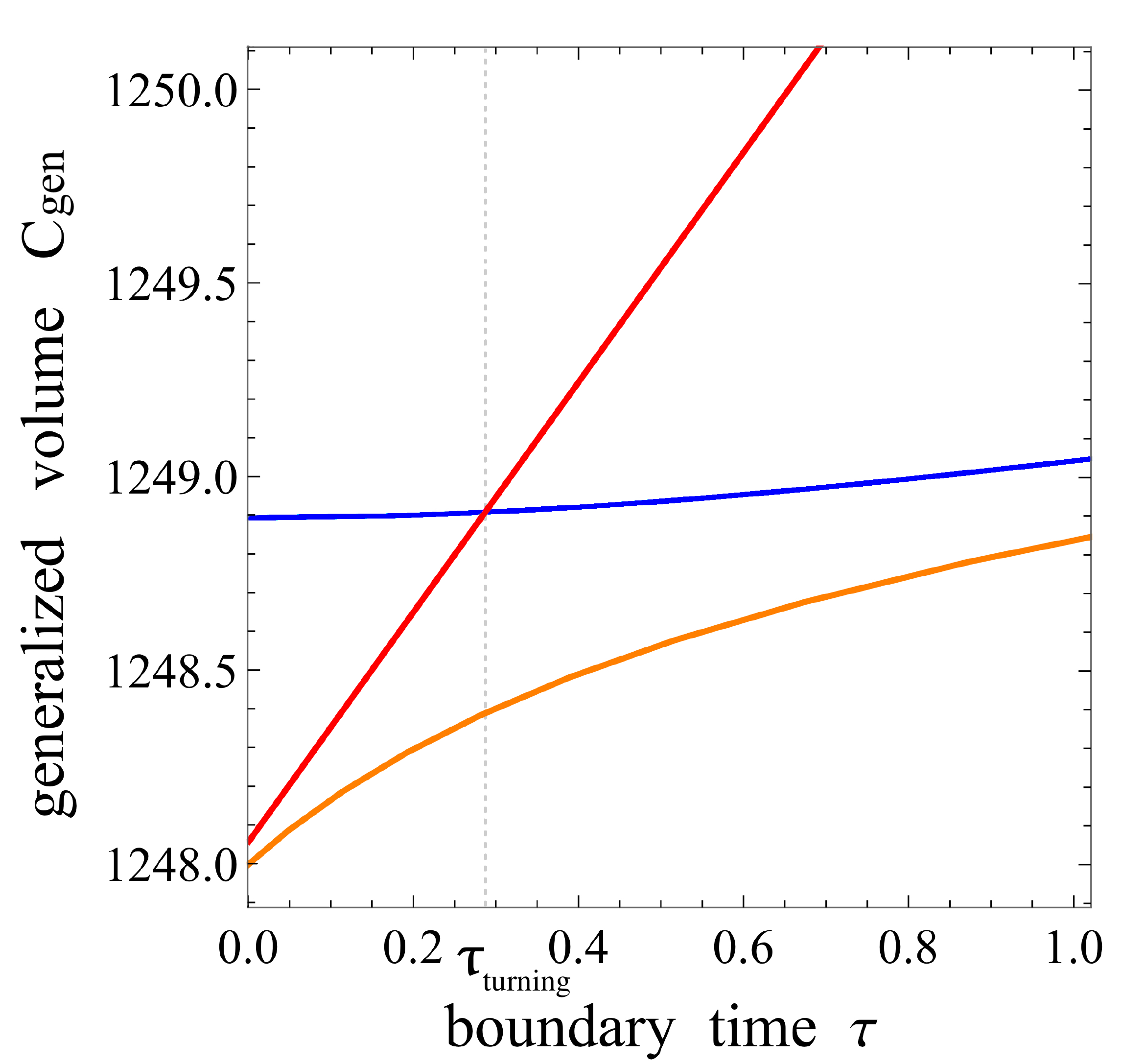}
  \caption{\label{cgent} The relation between the generalized volume-complexity$C_{\text{gen}}$ and the boundary time $\tau$ with $L/r_2=1$, $r_1/r_2=0.18$ and $\lambda =0.00025$. The UV cutoff $r_\epsilon/r_2 =500$. 
  The yellow curve ($C_{\text{genM}}$) represents the generalized volume-complexities of hypersurfaces corresponding to non-dipping branches.
  The blue curve ($C_{\text{genR}}$) represents the generalized volume-complexities of hypersurfaces corresponding to dipping branches with $P_v < P_{\text{crtR}}$.
  The red curve ($C_{\text{genL}}$) represents the generalized volume-complexities of hypersurfaces corresponding to dipping branches with $P_{\text{crtR}} < P_v < P_{\text{crtL}}$.
  }
 \end{figure}

As shown in Fig.~\ref{cgent}, the turning time $\tau_{turning}$ can be obtained. The yellow curve representing the generalized volume-complexity$C_{\text{genM}}$ corresponds to the dipping branch and has the minimum value at any given time.
The generalized volume-complexities of hypersurfaces corresponding to non-dipping branches compete with each other over time, and the branches representing the largest generalized volume-complexityalternate at the turning time.
 Before the turning time, the corresponding generalized volume-complexity$C_{\text{genL}}$ (the red curve) is smaller than $C_{\text{genR}}$ (the blue curve), and after the turning time,
the corresponding generalized volume-complexity$C_{\text{genL}}$ is larger than $C_{\text{genR}}$. This means that at the turning time $\tau_{\text{turning}}$, the hypersurface representing
the maximum generalized volume-complexityjumps from one branch to the other. That is, a hypersurface dual to the complexity of the thermofield double state defined on the boundary jumps from one branch to another.
This discontinuous jump is highly reminiscent of a phase transition, and the turning time denotes the moment at which this phase transition occurs.

Obviously, the turning time $\tau_{\text{turning}}$ depends on the shape of the effective potential $U(r)$. In other words,
it depends on the choice of the parameters $Q/M$, $L$, $\lambda$, and the definition of generalized volume-complexity$C_{\text{gen}}$.

\begin{figure}  
  \subfigure[~$\tau_{\text{turning}} - \lambda $]{\label{tl}
    \includegraphics[width=7cm]{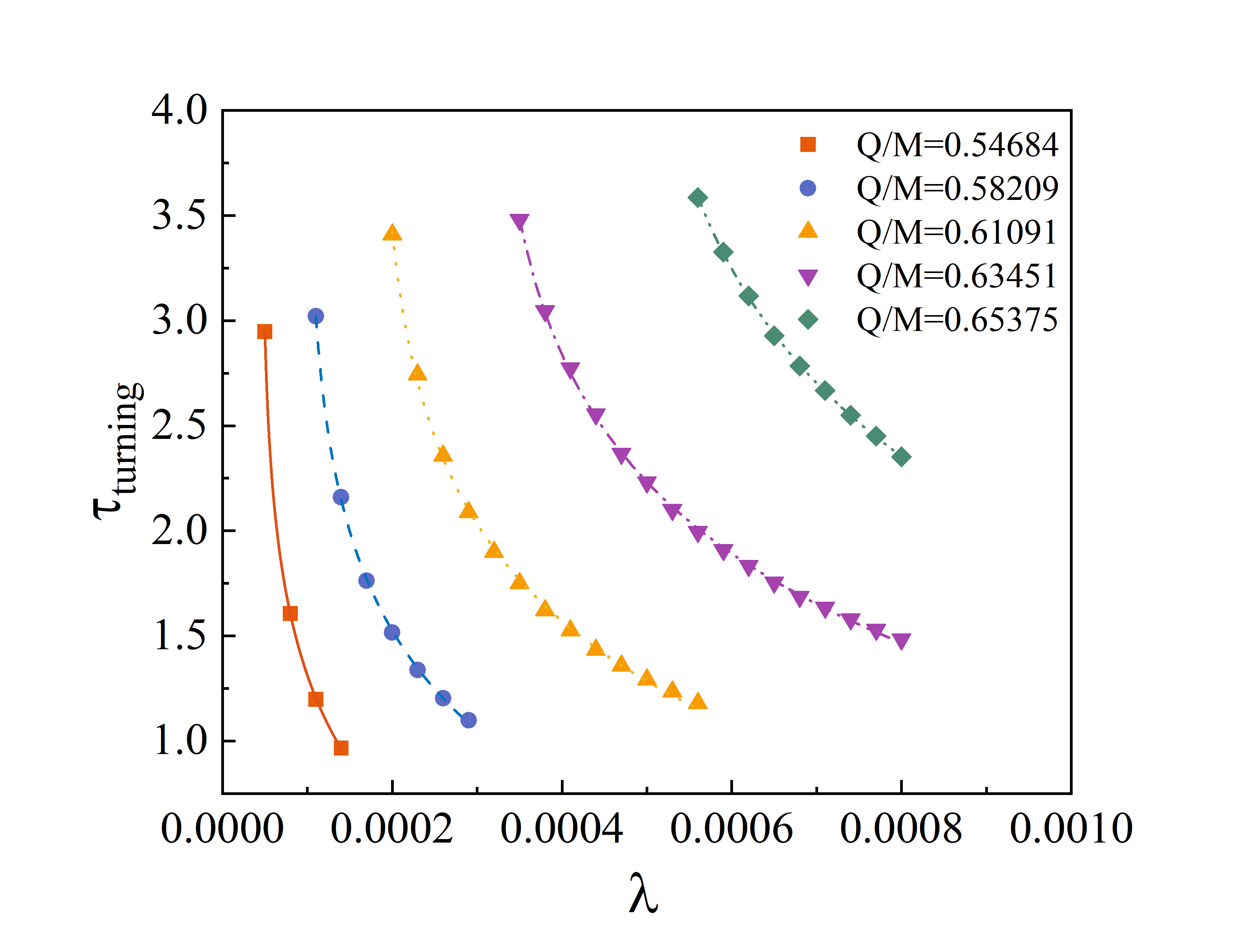}}
  \subfigure[~$\tau_{\text{turning}} -\ln (\lambda +b_3)$]{\label{tlnl}
    \includegraphics[width=7cm]{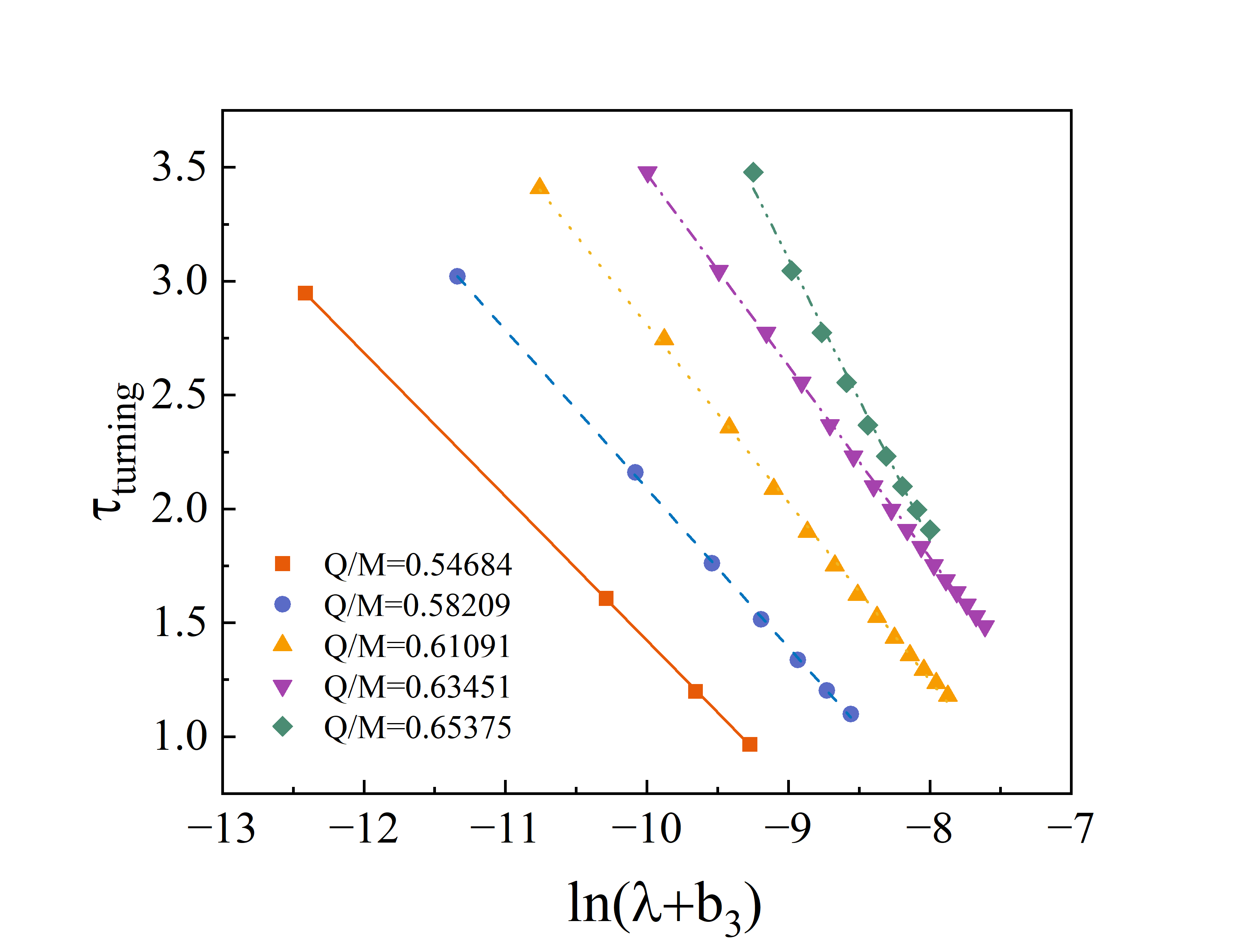}}
    \caption{Turning time $\tau_{\text{turning}}$ as a function of $\lambda $.}
  \end{figure}

We find that for the definition
\begin{equation}
  C_{\text{gen}}=2\frac{V_x}{G_N L}\int^{r_{\text{min}}}_{r_\epsilon }  \,dr \frac{r^{2(d-1)} a^2(r)} {\sqrt{P_{v}^{2}-U(r)}},
  \end{equation}
where $a(r)=1+\lambda L^4 C^2$, the turning time $\tau_{\text{turning}}$ decreases with the parameter $\lambda$ (see Fig.~\ref{tl}). As shown in Fig.~\ref{tlnl}, we observe a logarithmic relationship between the turning time $\tau_{\text{turning}}$ and the parameter $\lambda$ for a fixed $Q/M$:
\begin{equation}
  \tau =b_1 +b_2 \ln (\lambda +b_3),
  \end{equation}
where $b_1$, $b_2$ and $b_3$ are the fitting values shown in Table~\ref{tab1}.
As the turning time $\tau_{\text{turning}}$ increases, the late growth rate of the local maximum near the Cauchy horizon $P_{\text{crtL}}$ also increases.
This increase in the area growth rate leads to an earlier arrival of the turning time.
At the same time, with the increase of the parameter $\lambda$, the decreasing rate of the turning time $\tau_{\text{turning}}$ decreases.
Until the turning time reaches zero, the hypersurface corresponding to the non-dipping branch of the red curve will always have a larger generalized volume.

\begin{figure}  
  \subfigure[~$\tau_{\text{turning}} - Q/M $]{\label{tq}
    \includegraphics[width=7cm]{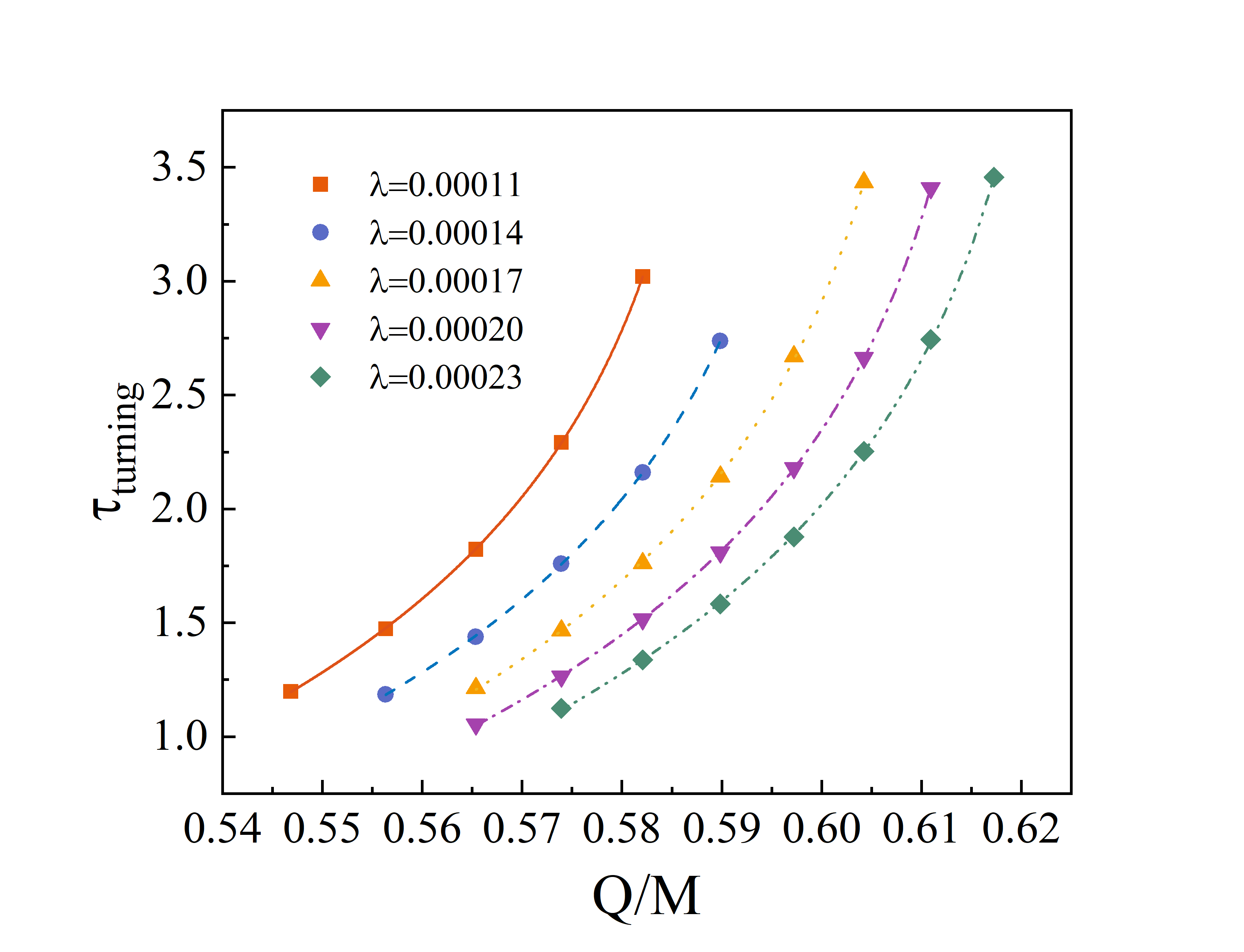}}
  \subfigure[~$\tau_{\text{turning}} -\ln (-Q/M +b'_3)$]{\label{tqln}
    \includegraphics[width=7cm]{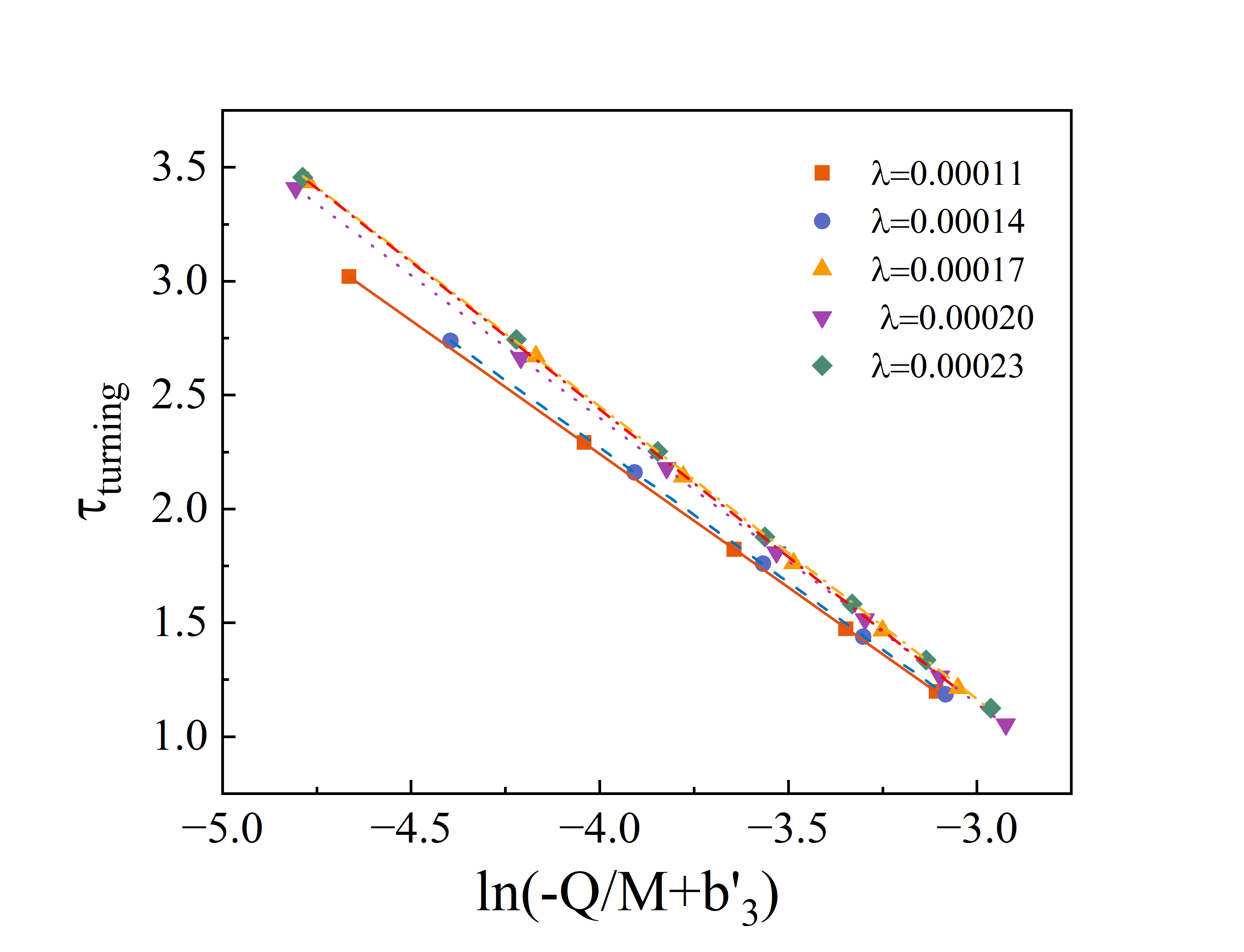}}
    \caption{Turning time $\tau_{\text{turning}}$ as a function of the charge-to-mass ratio $Q/M$.}
  \end{figure}

We have another conclusion: the turning time $\tau_{\text{turning}}$ increases with the charge-to-mass ratio $Q/M$, shown in Fig.~\ref{tq}. As shown in Fig.~\ref{tqln}, the relationship between the turning time $\tau_{\text{turning}}$ and the charge-to-mass ratio $Q/M$ closely follows a logarithmic function:
\begin{equation}
\tau = b'_1 + b'_2 \ln (-Q/M + b'_3),
\end{equation}
where $b'_1$, $b'_2$, and $b'_3$ are the fitting values shown in Table~\ref{tab2}.

Our findings propose a discontinuous variation in bulk physics that is
dual to the complexity of the thermofield double state defined on the boundary.
The turning time denotes the moment at which this discontinuous variation occurs.
We use the numerical method to fit the relationships of $\tau_{\text{turning}}\sim\lambda$ and $\tau_{\text{turning}}\sim Q/M$.
And in the parameter space we considered, both of them fit the logarithmic function well.

\begin{table}
  \centering
  \begin{tabular}{cccc}
    \hline\hline
    \multicolumn{4}{c}{$\tau =b_1 +b_2 \ln (\lambda +b_3)$}\\
    \hline
    $Q/M$& $b_1$& $b_2$& $b_3$\\
    \hline
    0.54684& -4.89634 & -0.63186& -4.59398E-5\\
    \hline
    0.58209& -4.87936& -0.69684& -9.80001E-5\\
    \hline
    0.61091& -4.828370& -0.75939& -1.80931E-4\\
    \hline
    0.63451& -4.71770& -0.81162& -3.09332E-4\\
    \hline
    0.65375& -5.54068& -0.98772& -4.63683E-4\\
    \hline\hline
  \end{tabular}
  \caption{\label{tab1} Turning time $\tau_{\text{turning}}$ as a function of $\lambda $, where $b_1$, $b_2$ and $b_3$ are the fitting values.}
\end{table}

\begin{table}
  \centering
  \begin{tabular}{cccc}
    \hline\hline
    \multicolumn{4}{c}{$\tau =b'_1 +b'_2 \ln (-Q/M +b'_3)$}\\
    \hline
    $\lambda $& $b'_1$& $b'_2$& $b'_3$\\
    \hline
    0.00011& -2.44840& -1.1724& 0.59151\\
    \hline
    0.00014& -2.46565& -1.18376& 0.60218\\
    \hline
    0.00017& -2.74669& -1.29589& 0.61269\\
    \hline
    0.00020& -2.61918& -1.25456& 0.61909\\
    \hline
    0.00023& -2.69168& -1.28528& 0.62560\\
    \hline\hline
  \end{tabular}
  \caption{\label{tab2} Turning time $\tau_{\text{turning}}$ as a function of the charge-to-mass ratio $Q/M$, where $b'_1$, $b'_2$, and $b'_3$ are the fitting values.}
\end{table}

\section{Discussion and conclusion}
\label{sec5}
In this paper, we studied the generalized volume-complexity in a 4-dimensional RN-AdS black hole.
Generalized volume-complexity is an extension of volume in the Complexity=Volume (CV) proposal, which was recently introduced
in Ref.~\cite{Belin:2021bga}. This quantity was defined by two scalar functions $F_1$ and $F_2$.
We chose the scalar functional as $F_1=F_2=1+\lambda L^4C^2$.
When $F_1=F_2=1$, it turns into the CV proposal.

We proved that for a 4-dimensional RN-AdS black hole, the generalized volume-complexity always has a linear growth rate $P_{\text{crt}}$ in the late period. The linear growth rate is determined by the maximum of the effective potential $U(r)$, which
is located between the two horizons. Furthermore, we found that the effective potential always has at least one local maximum between the two horizons, independent of the selection of the free parameters $\lambda$.
Additionally, there can be more than one local maximum between the two horizons for the certain parameter values.

That is, one can have multiple hypersurfaces, all of which take the same time to reach the boundary.
These new observables satisfy the characteristic of the thermofield double state, i.e., it grows linearly on the late stage.

The size relationship of the two generalized volume-complexities on the two hypersurfaces changes over time. This results in the substitution of the maximum extreme hypersurface which is dual to the complexity of the thermofield double state. We call the time when one hypersurface replaces another
to become the largest extreme hypersurface the turning time $\tau _{\text{turning}}$.
Our findings propose a discontinuous variation in bulk physics that is dual to the complexity of the thermofield double state defined on the boundary.
That is, a hypersurface dual to the complexity of the thermofield double state defined on the boundary jumps from one branch to another. For the field theory on the boundary, this shows that there is more than one evolutionary path from the reference state to the target state.
This discontinuous jump is highly reminiscent of a phase transition, and the
turning time denotes the moment at which this phase transition occurs. 
We found that the turning time $\tau_{\text{turning}}$ of the system increases with the increase
of the charge-to-mass ratio $Q/M$ and decreases with the increase of the free parameter $\lambda $.
In the range of the parameters we consider, they fit the logarithmic function well. 
It will be interesting to investigate whether this phase-like behavior is related to phase transitions recently explored in quantum many-body systems \cite{Roca-Jerat:2023mjs}.

  Finally, it should be pointed out that in addition to the examples given in the paper, we can extend the definition of the generalized function to
  \begin{align}
    a(r)=\sum_{j=0}^{n}(-1)^{j}\lambda_{j}(L^{4}C^{2})^{j},
\end{align}
where $\lambda_{j}$ are free parameters, $j$ and $n$ are integers. When $\lambda_{0}=1$ and $n=1$, the above generalized function can be reduced back to the case that we considered in this paper.
For these scalar functionals, the turning time can also be found in the Schwarzschild AdS black hole within certain parameter values. This suggests that our findings can be generalized.



%
%

%

\acknowledgments
  We would like to thank Le-Chen Qu and Shan-Ming Ruan for their very useful suggestions,
  comments, and sincere help. We also thank Jing Chen and Jun-Jie Wan for their helpful
  discussions. This work was supported by the National Natural Science Foundation of China (Grants No.~11875151 and No.~12247101), the 111 Project under (Grant No. B20063) and ``Lanzhou City's scientific research funding subsidy to Lanzhou University.



\end{document}